\documentclass[preprint,5p,times,twocolumn]{elsarticle}

\usepackage{amssymb}
\usepackage{caption}
\usepackage{subcaption}

\journal{arXiv}

\usepackage{outlines}
\usepackage{url}
\usepackage{todonotes}
\usepackage{float}
\usepackage{xcolor}
\usepackage{amsmath}

\begin{document}

\begin{frontmatter}



\title{Towards an automated workflow in materials science for combining multi-modal simulative and experimental information using data mining and large language models}

\cortext[cor1]{Corresponding author.}

\author[inst1,inst2]{Balduin Katzer\corref{cor1}}
\ead{balduin.katzer@kit.edu}
\author[inst1]{Steffen Klinder}
\author[inst1,inst2]{Katrin Schulz\corref{cor1}}
\ead{katrin.schulz@kit.edu}

\affiliation[inst1]{organization={Karlsruhe Institute of Technology (KIT), Institute for Applied Materials (IAM)},
            addressline={Kaiserstr. 12}, 
            city={Karlsruhe},
            postcode={76131}, 
            country={Germany}}
\affiliation[inst2]{organization={Karlsruhe University of Applied Sciences (HKA)},
            addressline={Moltkestr. 30}, 
            city={Karlsruhe},
            postcode={76133}, 
            country={Germany}}

\begin{abstract}
To retrieve and compare scientific data of simulations and experiments in materials science, data needs to be easily accessible and machine readable to qualify and quantify various materials science phenomena.
The recent progress in open science leverages the accessibility to data. However, a majority of information is encoded within scientific documents limiting the capability of finding suitable literature as well as material properties.
This manuscript showcases an automated workflow, which unravels the encoded information from scientific literature to a machine readable data structure of texts, figures, tables, equations and meta-data, using natural language processing and language as well as vision transformer models to generate a machine-readable database.
The machine-readable database can be enriched with local data, as e.g. unpublished or private material data, leading to knowledge synthesis.
The study shows that such an automated workflow accelerates information retrieval, proximate context detection and material property extraction from multi-modal input data exemplarily shown for the research field of microstructural analyses of face-centered cubic single crystals.
Ultimately, a Retrieval-Augmented Generation~(RAG) based Large Language Model~(LLM) enables a fast and efficient question answering chat bot.
\end{abstract}




\end{frontmatter}



\section{Introduction}
\label{sec:Introduction}
Understanding physical processes in materials and material microstructures is of fundamental importance in facilitating their use in engineering applications.
However, analyzing the increasing amount of existing scientific knowledge and extracting the relevant information for a desired research project is a challenging task.
Especially, combining information from experiments, simulations and theory is of great significance as different aspects are considered at each discipline that together, ultimately, form a holistic picture~\cite{Pollock2013a, Pablo2014, Wei2019, Choudhary2022a}.
Machine learning~(ML) and artificial intelligence~(AI) have been recently used as advanced computational tools to accelerate the physical understanding in materials science research~\cite{Wei2019, Morgan_2020, Guo_2021, Choudhary2022a, Jain2024}.
Recent progress in these computational methods enabled AI-assisted models with the ability to extrapolate beyond their data basis and generate novel materials science approaches, called generative AI~(genAI)~\cite{Fuhr_2022, Wang2023}.
Applying genAI leads for example to a novel design of crystalline materials~\cite{Zhao2023a}, of molecule properties~\cite{Manica2023} and of architected materials~\cite{Lew2023}.
A fundamental deep learning architecture of many genAI models is the transformer architecture, which possesses a self-attention mechanism leading to contextual awareness of data~\cite{Vaswani2017}.
This transformer model is the foundation of the Large Language Model~(LLM), which is a context-aware genAI model for natural language processing~(NLP) such as Generative Pre-Trained Transformer~(GPT)~\cite{Lei_2024, Yenduri_2024}.
The performance of transformer models are particularly characterized by the quality and the amount of data for pre-training leading to more powerful LLMs over the past years~\cite{Yenduri_2024}.
Based on this progress, materials science research has become more accessible due to the sole use of natural language input.
For example, in additive manufacturing, novel material designs and entire manufacturing processes are derived by LLMs~\cite{Hsu_2022, Chandrasekhar_2024}.
A variety of other example usages for LLMs in materials science are showcased in a study of Jablonka et al.~\cite{Jablonka_2023} including knowledge discovery, property prediction as well as advances in user-model interfaces.

To enable a more accelerated and tailored investigation to a domain-specific research area, LLMs are seen to have great potential.
In particular, two main strategies have been developed for LLMs for this task, fine-tuning and Retrieval-Augmented Generation~(RAG)~\cite{Yu2024}.
Fine-tuning of a LLM bases on additional model training with domain-specific data fitting the model parameters to its specific task.
This procedure leads to several mechanics and materials science LLMs such as MechGPT~\cite{Buehler_2024} and ProtAgent~\cite{Ghafarollahi_2024} or to scientific LLMs in general like SciBERT~\cite{Beltagy_2019}.
In contrast, RAG is a method to retrieve the relevant information from a user-specific database without modifying the transformer model itself, while being more context-aware but less generalizable~\cite{Lewis2020}.
For utilizing RAG, various machine-readable databases could be considered from which information is retrieved. However, natural language databases are employed most frequently~\cite{Foppiano_2024, Prince2024}.

\begin{figure}[t]
    \centering
    \includegraphics[width=\linewidth]{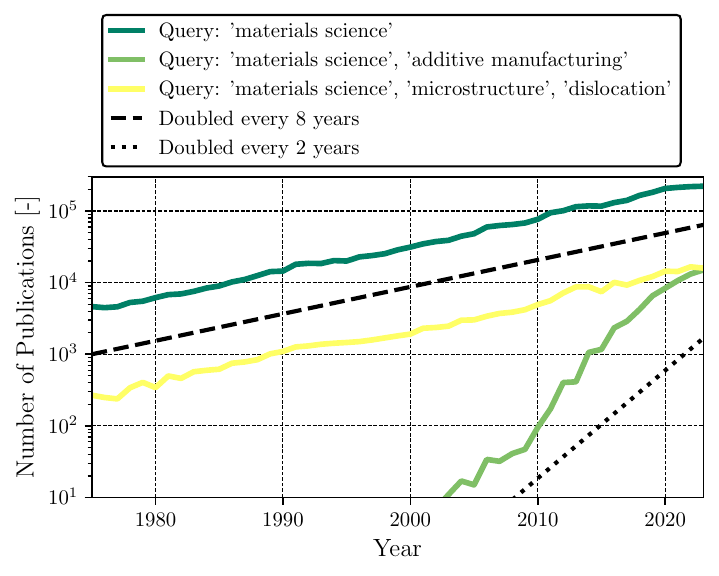}
    \caption{Number of publications in material science during the past 50 years for different queries based on data from dimension.ai~\cite{appdimensions}.}
    \label{fig:publications}
\end{figure}

Generating such a database of high relevance for a specific research area is challenging and tedious.
The data can originate from various sources such as scientific literature or user specific data.
As an example, the highly increasing amount of scientific literature in materials science is depicted in Fig.~\ref{fig:publications}.
It shows the number of scientific publications for different keyword queries based on the dimension.ai~\cite{appdimensions} database within the last 50 years.
A general trend in scientific research can be deducted, e.g., the number of scientific literature doubles every eight years for materials science in general, whereas it doubles every two years for research in additive manufacturing in materials science indicated by the dashed and dotted lines.
To incorporate the increasing amount of data, workflows have been developed to automatize database generations from scientific literature~\cite{Gilligan2023} and LLMs have been used to extract accurate information from documents~\cite{Polak2024}.
For example, a RAG based LLM is utilized for additive manufacturing to answer user-defined questions from literature data~\cite{Chandrasekhar_2024}, or in another example, user-specific data in the form of electronic lab notebooks is utilized for question-answering in a LLM workflow~\cite{Jalali2024}.
However, most of the previous approaches focus purely on databases generated from text data without using the richness of multi-modal data sources.
But recent progress in transformer model approaches lead to the generation of a Large Multi-modal Model~(LMM), which is capable of contextualizing multi-modal input such as audio or images~\cite{Lei_2024, Buehler_2024c, Buehler2024d, Picard2023}.
Thus, workflows are required, which generate multi-modal databases and are capable of processing multi-modal data within the LLM.

This study proposes an automated workflow, which incorporates multi-modal data from scientific literature as well as multi-modal user-specific data to deduct a multi-modal database, which is subsequently utilized for a RAG based LLM for question-answer prompting.
This work addresses the challenge of automatically and accurately identifying the most suitable scientific research related to the research domain of dislocation-microstructure based materials science.
In this domain, different length scales need to be covered by researchers' investigations leading to various materials science theories, each evaluated by a multitude of simulative and experimental methods~\cite{Bertin_2020}.
This workflow tackles the search for simulations, experiments, or theories that can be taken into account to support and compare user-defined research questions.
This study attempts to aid the researcher's need for a quick and accurate retrieval of suitable research data. The following key research questions are addressed:
\begin{enumerate}
    \item To what extent is a data-driven workflow and a domain-specific RAG based LLM able to detect and represent most suitable scientific features based on user-specific queries?
    \item What are the current limitations of the automatized multi-modal workflow, i.e., to what extent does the domain-specific RAG based LLM reply accurate results and what is the reason for inaccurate results?
\end{enumerate}

The manuscript is structured as follows: 
Section~\ref{sec:Methods} introduces the data mining tools, the user-specific data as well as the transformer models.
Section~\ref{sec:Results} displays the results of the workflow.
The current limitations as well as the accuracy of the results are discussed in Section~\ref{sec:Discussion}.
Section~\ref{sec:Conclusion} gives a summary and outlook of the automated workflow and the RAG based LLM approach.

\begin{figure*}[h]
    \centering
    \includegraphics[width=\textwidth]{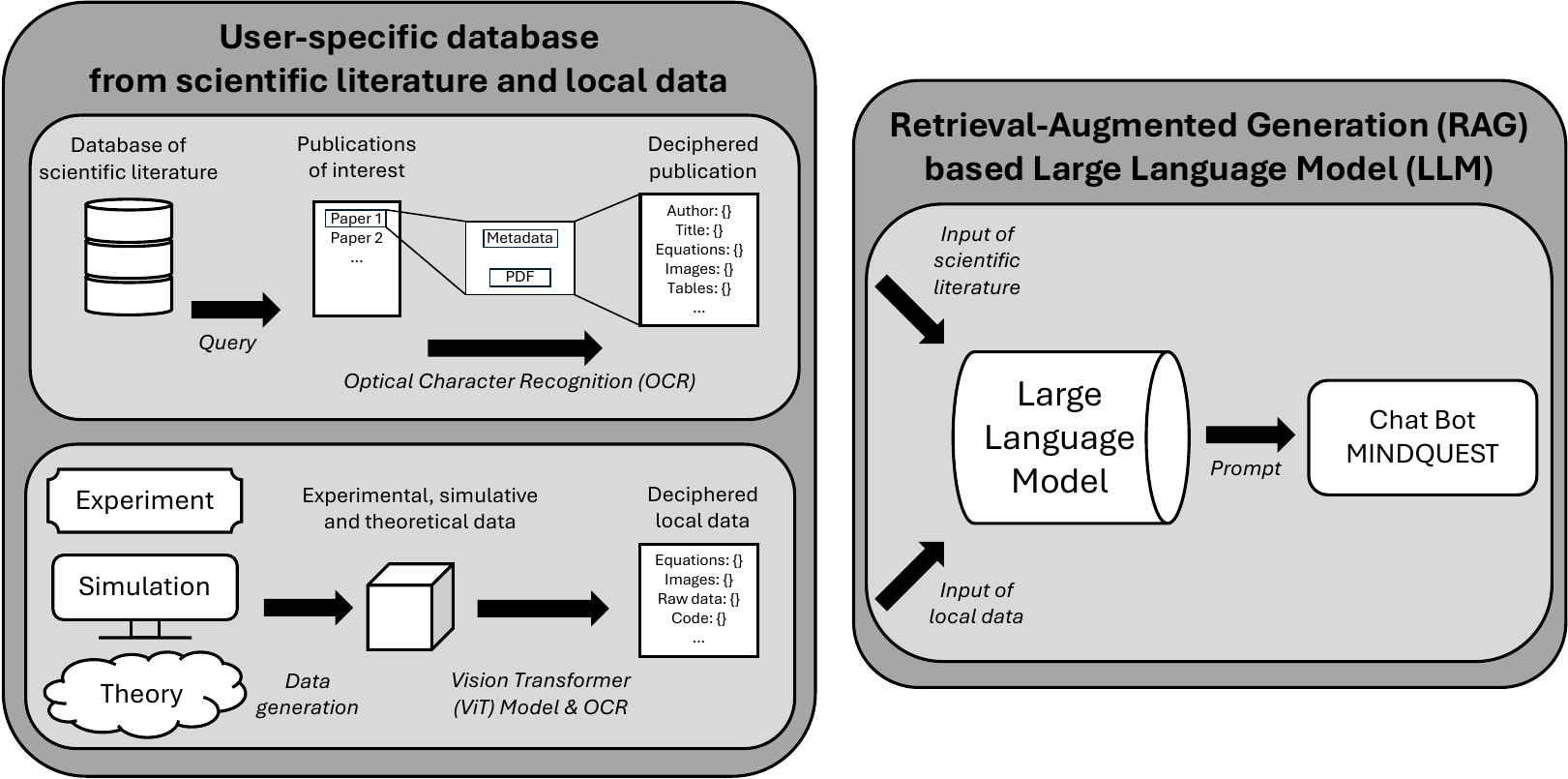}
    \caption{Automated workflow to generate a Retrieval-Augmented Generation~(RAG) based Large Language Model~(LLM) using a multi-modal database.}
    \label{fig:workflow}
\end{figure*}
%

\section{Methods}
\label{sec:Methods}
This study introduces a method, which enables an automated workflow to query and process literature in the domain of materials science 
in combination with local user-specific data from experiments, simulations or theory. Using a Retrieval-Augmented Generation~(RAG) based Large Language Model~(LLM), the objective is to provide a faster and more accurate retrieval of information.
The schematic of the workflow is shown in Fig.~\ref{fig:workflow}.
The top left box represents the collection of relevant scientific literature resulting from conducting a query for a desired research question within a literature database.
This query identifies possible publication candidates based on keyword matching.
Subsequently, full-text documents for each candidate of interest are deciphered and structured into various document entities like texts, equations, images, tables, and meta-data including, e.g., the authors, the title, or the doi.
A transformation is performed through a pipeline of machine learning models that includes layout detection, data cleaning, and optical character recognition~(OCR).
In Fig.~\ref{fig:workflow}, the bottom left box illustrates the process of generating a structured database from the local and user-specific experimental, simulative, or theoretical data.
In the next step, the structured data from the literature and the local data from the user are combined into a user-specific database.
Ultimately, an LLM chat bot is created which retrieves information from the combined database to answer user-specific questions by taking into account literature as well as local information and results.
The automated workflow is exemplarily established in the following for the field of dislocation-microstructure based materials science. 

\subsection{Data mining from literature via OCR models}
\label{subsec:datamining}
Most scientific literature is provided to its community by PDF documents.
In materials science, each document is rich in information including multi-modal information in various forms such as texts, equations, figures or tables.
However, the machine readability of PDF documents is limited.
Thus, each document needs to be deciphered into a machine-readable structured dataset to provide better accessibility.
Recent progress in OCR models enable the transformation of PDF documents into machine-readable markup language incorporating tables, equations and images.
Table~\ref{tab:OCR} shows the comparison of the OCR models \textit{pypdfium2}\footnote{https://github.com/pypdfium2-team/pypdfium2}, \textit{nougat}\footnote{https://facebookresearch.github.io/nougat}~\cite{Blecher2023} and \textit{marker}\footnote{https://github.com/VikParuchuri/marker} with respect to their capability to properly extract equations, images, tables and text.
\textit{pypdfium2} is a fast OCR model, however, unable to correctly depict equations and prone to errors in general.
\textit{nougat} and \textit{marker} are more advanced machine learning OCR models, which are able to convert PDF documents incorporating equations and tables with high precision, however, with high computational cost.
Regarding the field of application, \textit{marker} is considered most adequate for incorporation into an automated workflow since it is able to extract images of the PDF document by incorporating the layout detection tool \textit{surya}.
Here, \textit{marker} detects each image within each document leading to a set of image data in addition to the markdown file for each manuscript. 
This is, e.g., particularly important for the incorporation of experimental or simulation results often presented as images or diagrams.
An interface of \textit{marker} can then be used to couple it again with \textit{pypdfium2} and \textit{nougat} to harmonize accuracy and speed of the OCR.

\begin{table}[h]
    \centering
    \caption{Comparison of multi-modal OCR capabilities across three different models (\textit{pypdfium2}, \textit{nougat}, \textit{marker}) to recognize and process different types of content in documents. \textit{marker} shows the broadest capabilities, handling text, tables, equations, and images, while the other models have limitations.
    }
    \begin{tabular}{ccccc}
         \hline
         OCR model & Text & Tables & Equations & Images \\
         \hline
         \textit{pypdfium2} & yes & yes & no & no\\
         \textit{nougat} & yes & yes & yes & no\\
         \textit{marker} & yes & yes & yes & yes\\
         \hline
    \end{tabular}
    \label{tab:OCR}
\end{table}

\subsection{User-specific microstructure simulation data}
\label{subsec:SimulationData}
\begin{figure}[h]
    \includegraphics[width=\linewidth]{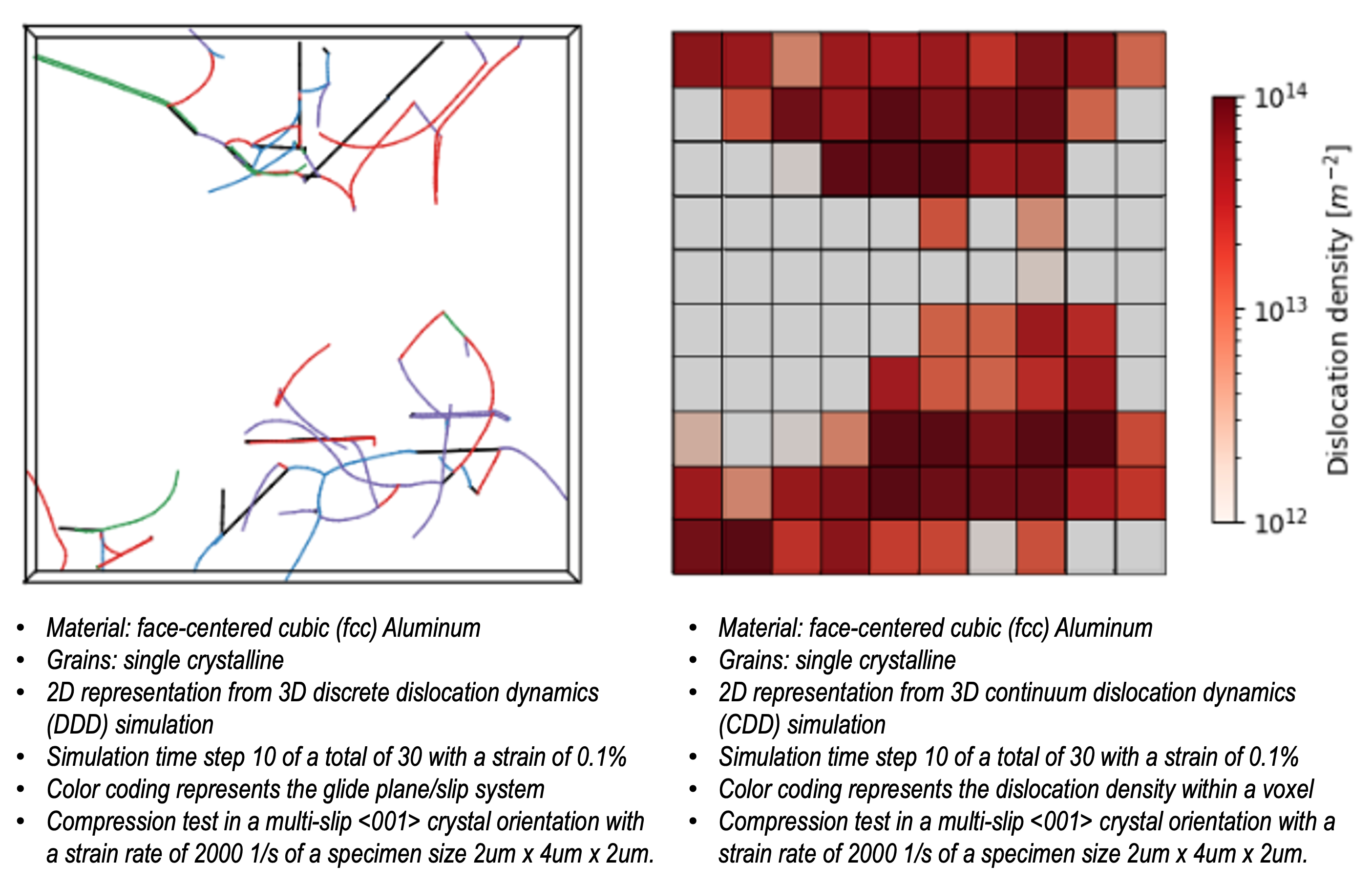}
    \caption{Example images of a DDD and a CDD microstructure (given as 2d slice of a 3d material system) including metadata information about the simulation features.}
    \label{fig:DDDCDD}
\end{figure}

The present workflow aims to incorporate local user-specific data into the retrieval database.
This study exemplarily examines local data from microstructure simulations of single-crystalline aluminum employing two different simulation approaches that consider different length scales for the resolution of microstructural defect structures.
The Discrete Dislocation Dynamics~(DDD) approach resolves the dynamics of the evolution of the dislocation microstructure during plastic deformation at a discrete level showing individual atomistic defects~\cite{Weygand2001, Katzer2022, Lee2023, Katzer2024}.
The Continuum Dislocation Dynamics~(CDD) approach models the microstructure evolution in a homogenized form by using continuum fields for the dislocation density~\cite{Schulz2019, Sudmanns_2020, Zoller_2021, Katzer2024b}.
For this study, 2D~images are generated from three-dimensional DDD and CDD data.
A set of example images are depicted in Fig.~\ref{fig:DDDCDD}.
Additionally, each dislocation microstructure image is enriched by a set of features providing additional data for image interpretation and contextualization, e.g., including information about the material, the simulation set-up or the strain state.

\begin{figure*}[h]
    \centering
    \includegraphics[width=\linewidth]{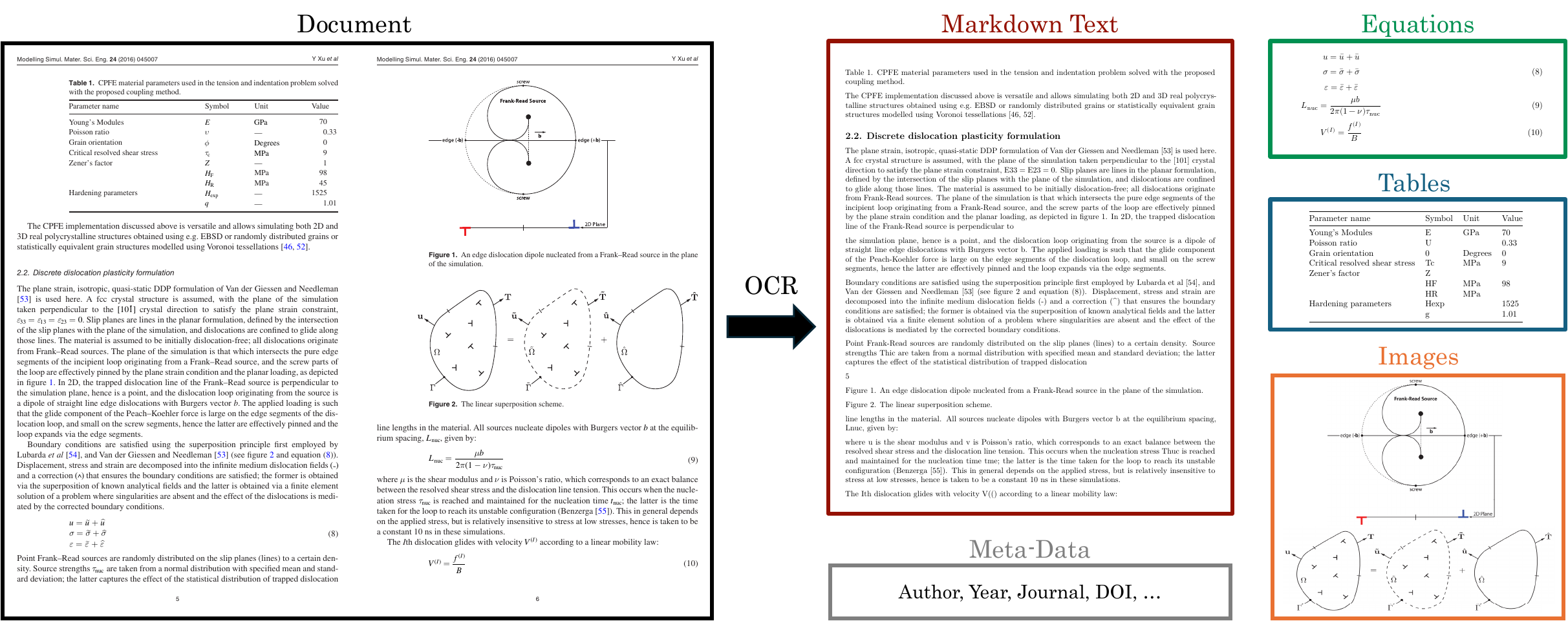}
    \caption{
    Example for deciphering full-text PDF document into structured database entities.
    Each document is subdivided into the following entities: markdown text, meta-data, equations, tables and images.
    (here, \cite{Xu_2016} reproduced with permission from IOP Publishing under a Creative Commons License)
    }
    \label{fig:structure}
\end{figure*}

\subsection{Transformer models}
\label{sub:Transformer}
Transformer models capture contextual relationships within data, e.g., for natural language tasks as well as for visual tasks.
The models utilize embeddings to map semantic and visual information into a latent space. From this latent space, the proximity between the data can be retrieved.
The proposed workflow considers three types of transformer models: Embeddings models~(EM), large language models~(LLM), and vision transformer~(ViT) models.

The embedding model identifies the relevant information from a fragment of text and generates a $n$-dimensional embedding vector.
This approach employs the \textit{all-minilm} model with 22B parameters, which is a light-weight and fast embedding model, and based on the Bidirectional Encoder Representations from Transformers~(BERT)~architecture~\cite{Wang2020}.
The large language model is able to generate text from user-specific queries by learning statistical relationships of natural language data.
Here, the foundation language model from Meta AI~\textit{LLaMA}~3.2 with 3B parameters is considered.
\textit{LLaMA} is chosen due to its high performance and its precise contextual analysis compared to other LLM models such as BERT~\cite{Touvron2023}.
The vision transformer model is able to generate text based on images by generating an embedding vector from patches of an image.
In this study, the Large Language and Vision Assistant model~\textit{LLaVA} with 7B parameters is considered~\cite{Liu2023}. The model is applied by the \textit{ollama}\footnote{https://ollama.com (Oct 2024)} framework.
Furthermore, OpenAI's generative pre-trained transformer \textit{ChatGPT-4o}\footnote{https://chatgpt.com (Oct 2024)} is used for comparison between the considered LLM and ViT models due to its ability to directly transform PDF documents into text.

\section{Results}
\label{sec:Results}

\subsection{Generation of a multi-modal machine-readable database}
\label{subsec:CorrectOCR}

The first step towards an automated workflow is the generation of a multi-modal machine-readable database for experimental and simulative materials science data.
Thus, the main objective is deciphering any input data of interest such as scientific literature or user-specific data to a given machine-readable output format.

\subsubsection{Finding relevant literature candidates}
To find the relevant scientific literature that applies to a certain research topic, a keyword-based initial screening is carried out to find the relevant publication candidates.
Therefore, the \textit{Semantic Scholar API}\footnote{https://www.semanticscholar.org/product/api (Oct 2024)} is used, which searches for keyword matches in the metadata, the title, and the abstract.
This study focuses on the retrieval of the most suitable scientific literature for dislocation-based plasticity of single crystalline face-centered cubic~(fcc)~materials.
Thus, the following keywords are considered for the screening query in the Semantic Scholar database: \verb|"dislocation"|, \verb|"plasticity"|, \verb|("face-centered cubic" or "fcc")| and \verb|"single crystal"|.
This query yields $\approx$2000 scientific publication candidates.
Due to license limitations and the lack of digitization of the earlier literature, this query led to $\approx$1200 full-text PDF documents, which is the corpus of the domain-specific literature of all subsequent analyses.

\begin{table}[]
    \centering
    \caption{
    Regular expression operators for identifying document entities including headings, paragraphs, equations, tables, figures, codes or citations, which enables further structuring of the markdown data into more distinct constituents.}
    \begin{tabular}{l l}
        Document entity & Regular expression operators\\\hline
        Heading & \verb|^#{1,6} .*| \\
        Paragraph & \verb|(.*\n)+| \\
        Equation & \verb|$.*?$| (inline) and \verb|$$.*?$$| (block) \\
        Table & \verb|^\.*?$| \\
        Figure & \verb|!\[.*?\]\(.*?\)| \\
        Code Block & \verb|`.*?`| (inline) or \verb|```.*?```| (block) \\
        Citation & \verb|\[.*?\]\(.*?\)|
    \end{tabular}
    \label{tab:regex}
\end{table}
\begin{figure*}[h]
    \begin{minipage}[b][][b]{0.35\textwidth}
        \includegraphics[width=\linewidth]{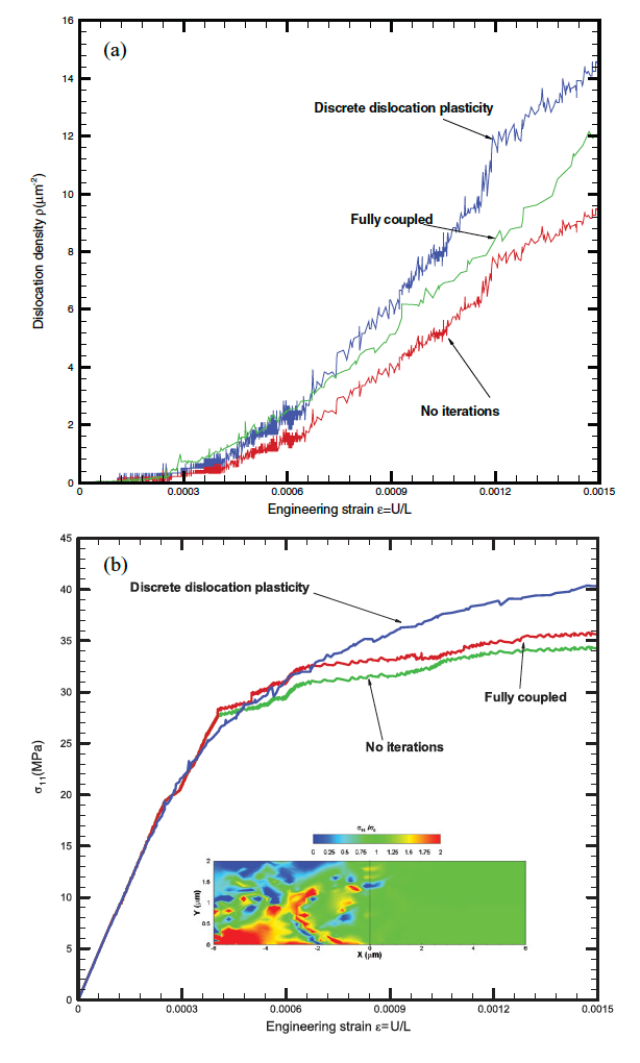}
    \caption*{(a) Image}
    \end{minipage}
    \hfill
    \begin{minipage}[b][][b]{0.3\textwidth}
        \tiny
        \itshape
        The image contains two graphs, labeled (a) and (b), which describe material behavior under mechanical deformation. Graph (a) shows the relationship between dislocation density and engineering strain, while Graph (b) depicts stress versus engineering strain. 
        
        In Graph (a), the x-axis represents engineering strain (\( \varepsilon = U/L \)), ranging from 0 to 0.0015, and the y-axis represents dislocation density (\( \rho \)), measured in \(\mu\text{m}^{-2}\), ranging from 0 to 16. Three curves are displayed, corresponding to different models: ``Discrete dislocation plasticity'' (blue), ``Fully coupled'' (green), and ``No iterations'' (red). The blue curve exhibits the highest dislocation density across all strains, with several rapid increases at higher strains, suggesting active dislocation generation. The green curve shows moderate growth, while the red curve represents the lowest dislocation density, indicating limited dislocation activity. These results suggest that the discrete dislocation plasticity model captures a more detailed and realistic depiction of dislocation interactions under strain, while the other models simplify the mechanical response.
        
        Graph (b) plots stress (\( \sigma_{11} \), in MPa) on the y-axis, ranging from 0 to 45 MPa, against engineering strain (\( \varepsilon \)), which is again plotted on the x-axis over the range 0 to 0.0015. The stress-strain response is shown for the same three models. The blue curve (discrete dislocation plasticity) exhibits the highest stress values across all strains, indicating a more robust response. The green curve (fully coupled) shows slightly lower stresses, while the red curve (no iterations) presents the lowest stress response, indicating a simplified mechanical model. At small strains (\( \varepsilon < 0.0005 \)), the differences between the models are minimal, but they diverge significantly as strain increases. The inset diagram in Graph (b) illustrates the spatial distribution of normalized resolved shear stress (\( \tau_{xy}/\tau_y \)) in a microstructure. The color map ranges from blue (\( \tau_{xy}/\tau_y = 0 \)) to red (\( \tau_{xy}/\tau_y = 2 \)), highlighting regions of high shear stress, which are potential sites for intense dislocation activity or failure initiation.
        
        Overall, the discrete dislocation plasticity model demonstrates superior performance in capturing both dislocation dynamics (Graph a) and the stress response (Graph b), making it more representative of real material behavior. The fully coupled and no iterations models simplify the mechanical response, resulting in lower stress and dislocation densities. The graphs highlight the importance of incorporating dislocation-level details for accurate material modeling, especially at higher strains where plastic deformation mechanisms dominate. Abrupt increases in dislocation density for the discrete dislocation plasticity model at higher strains, as seen in Graph (a), may indicate strain-hardening effects or sudden activation of dislocation sources. Additionally, the inset in Graph (b) underscores the significance of localized stress concentrations in understanding failure mechanisms.
    \caption*{(b) \textit{ChatGPT-4o}}
    \end{minipage}
    \hfill
    \begin{minipage}[b][][b]{0.3\textwidth}
        \tiny
        \itshape 
        The image consists of two parts: (a) and (b).

        Part (a) is a graph that shows the dislocation density as a function of engineering strain for different simulation scenarios. The x-axis represents the engineering strain ($\varepsilon = U/L$), and the y-axis represents the dislocation density (in $\mu\text{m}^{-2}$). There are three curves in the graph, each representing a different simulation scenario: ``Discrete dislocation plasticity'' (blue curve), ``Fully coupled'' (red curve), and ``No iterations'' (green curve). The ``Discrete dislocation plasticity'' curve shows a fluctuating increase in dislocation density as the strain increases, indicating the formation and movement of dislocations within the material. The ``Fully coupled'' curve also shows an increase in dislocation density but with a smoother trend, suggesting a more stable simulation model. The ``No iterations'' curve shows a relatively flat trend, indicating that no dislocations are formed or moved in this scenario.
        
        Part (b) is another graph that shows the stress ($\sigma$, in MPa) as a function of engineering strain for the same three simulation scenarios. The x-axis represents the engineering strain ($\varepsilon = U/L$), and the y-axis represents the stress (in MPa). The graph also includes a color-coded map in the bottom left corner, which represents the distribution of the stress component $\sigma_{11}$ (in MPa) across the material. The color scale ranges from blue (low stress) to red (high stress). The ``Discrete dislocation plasticity'' curve shows a sharp increase in stress as the strain increases, indicating the onset of plastic deformation. The ``Fully coupled'' curve shows a similar trend but with a slightly lower stress value, suggesting a more accurate simulation model. The ``No iterations'' curve shows a relatively flat trend, indicating that no stress is developed in this scenario.
        
        The scientific context of the image is related to the study of plastic deformation mechanisms in materials. Dislocation density and stress-strain behavior are key indicators of the material's ability to deform plastically. The ``Discrete dislocation plasticity'' scenario represents a simulation model that takes into account the formation and movement of dislocations within the material, which is a common mechanism of plastic deformation. The ``Fully coupled'' scenario represents a more advanced simulation model that couples the dislocation dynamics with the stress-strain behavior, providing a more accurate representation of the material's behavior under load. The ``No iterations'' scenario represents a simulation model that does not take into account the dislocation dynamics, resulting in a simplified and less accurate representation of the material's behavior.
        
        In summary, the image provides a detailed comparison of the dislocation density and stress-strain behavior of a material under different simulation scenarios. The ``Discrete dislocation plasticity'' scenario shows a more realistic representation of the material's behavior, with a fluctuating increase in dislocation density and a sharp increase in stress. The ``Fully coupled'' scenario shows a more accurate representation of the material's behavior, with a smoother increase in dislocation density and a slightly lower stress value. The ``No iterations'' scenario shows a simplified and less accurate representation of the material's behavior, with a relatively flat trend in both dislocation density and stress. The color-coded map in part (b) provides a visual representation of the stress distribution across the material, with red indicating high stress and blue indicating low stress.
    \caption*{(c) \textit{LLaVA}}
    \end{minipage}
    \caption{(a) 
    Example of a figure description using an image from \cite{Xu_2016} (reproduced with permission from IOP Publishing under a Creative Commons License) with the corresponding image description by the (b) \textit{ChatGPT-4o} and (c) \textit{LLaVA} Vision Transformer model.
    Both models demonstrate their capability to process and interpret visual input into accurate textual descriptions enabling the substitution of images with the corresponding textual representations.
    }
    \label{fig:exampleVT}
\end{figure*}

\subsubsection{Example of machine-readable data (re-)construction}
The considered corpus of scientific literature consists of full-text PDF documents.
The generation of a machine-readable database from these documents is performed applying the OCR models introduced in Section~\ref{subsec:datamining}.
Applying \textit{marker} leads to a markdown file and a set of images for each PDF document.
An example of a layout analysis of parts of a PDF document consisting of text, equations, tables, and figures is provided in~\ref{appendix:layout}.
This example demonstrates how the OCR model accurately deciphers the PDF document into various document entities.
Based on the generated machine-readable markdown file, the data is further subdivided into equations, tables, texts and metadata as shown in Fig.~\ref{fig:structure}.
The entities of the document are generated by regular expression operations depicted in Tab.~\ref{tab:regex} such as headings, equations, figures, or code blocks. This keeps the structure of the document in a logical form.

\subsubsection{Text generation from images via visual transformers}
Vision transformer models such as \textit{ChatGPT-4o} and \textit{LLaVA} as introduced in Section~\ref{sub:Transformer} enable the transformation of images into descriptive text. 
In materials science, images are key for many fundamental research results.
They are a significant part of the literature corpus as well as of local user-specific data, i.e. microstructural simulation results in this study.
The applied models identify and decode the visual features of an image into a natural language description.
The role and prompt considered by the ViT model for the transformation of each image is shown in~\ref{appendix:VT}.

An example of the transformation of a domain-specific image into natural language is shown in Fig.~\ref{fig:exampleVT}.
It illustrates the comparison of the natural language description given by \textit{ChatGPT-4o} and \textit{LLaVA}.
It appears that both models describe the image in reasonable detail, including indexing, labeling, axis annotation, and graph description.
This leads to the conclusion that both ViT models demonstrate the capability to accurately describe images related to the domain of materials science as text.
Based on the fact that only \textit{LLaVA} is open source, it has been integrated into the present workflow.
Subsequently, in this study, each image - from the publication database as well as local data - is transformed into a natural language description and integrated into the markdown file.

\subsubsection{Storage in vector database}
To efficiently store and retrieve textual data, a vector database is utilized in conjunction with a BERT-based embedding model as introduced in Section~\ref{sub:Transformer}.
Therefore, each markdown document is split into chunks of text of $n_{chunk} = 500$ characters.
For each text chunk, a numeric vector embedding is generated by using the \textit{all-minilm} embedding model.
Additionally, a document ID for each document and a chunk ID for each text chunk are created.
This ensures a unique, reproducible ID for each text chunk and each document, enabling consistent information retrieval.
Subsequently, the text chunks, the IDs as well as the embeddings are stored in a vector database.
This setup enables semantic search to query the database with natural language inputs and retrieve relevant results based on an embedding vector similarity metric.
This study employs the squared euclidean distance~(L$_2$)
\begin{equation}
    d = \sum_{i=1}^{n} (A_i - B_i)^2.
\end{equation}
Ultimately, the vector database in combination with the embedding vectors leverages the search for the most suitable text chunks with respect to a envisaged query.

\begin{figure}[t]
    \centering
    \includegraphics[width=\linewidth]{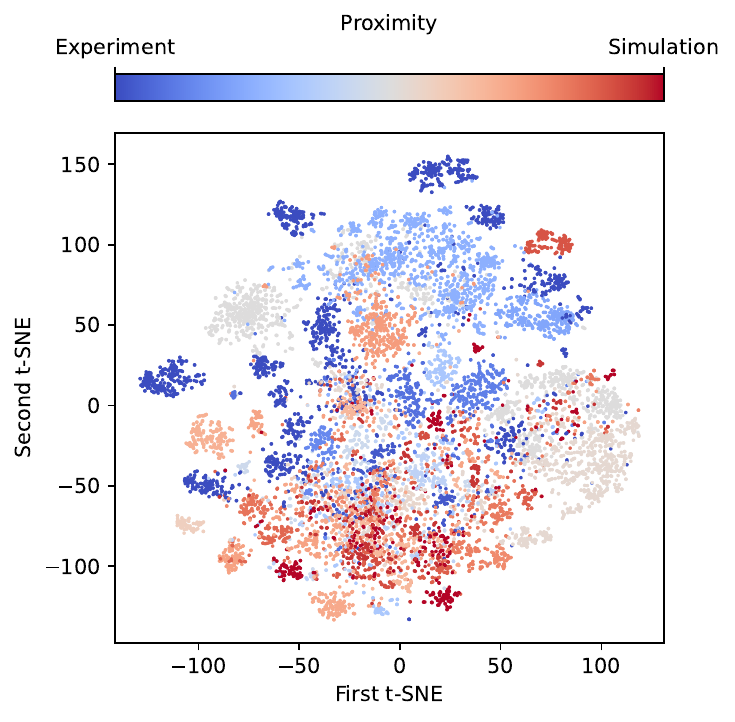}
    \caption{The t-SNE landscape of the scientific literature with colors indicating the proximity of each text chunk to either experimental or simulation-related research.}
    \label{fig:Landscape}
\end{figure}

Fig.~\ref{fig:Landscape} shows results for the landscape of the corpus of considered literature with respect to different domain-specific features.
A plot of the two-dimensional t-distributed stochastic neighbor embedding~(t-SNE) is depicted, where individual text chunks are mapped in a two-dimensional space.
Each text chunk is color-coded to indicate its proximity to experimental or simulation setups. The clustering of simulation and experimental data in the t-SNE landscape illustrates the proximity provided by the textual representation.

\subsection{A RAG based LLM}
\label{subsec:LLM}
In this work, a RAG based LLM is employed by retrieving the most similar data from the vector database introduced in Section~\ref{subsec:CorrectOCR} and by employing a question-answer chat bot.

\begin{figure}[t]
    \centering
    \includegraphics[width=\linewidth]{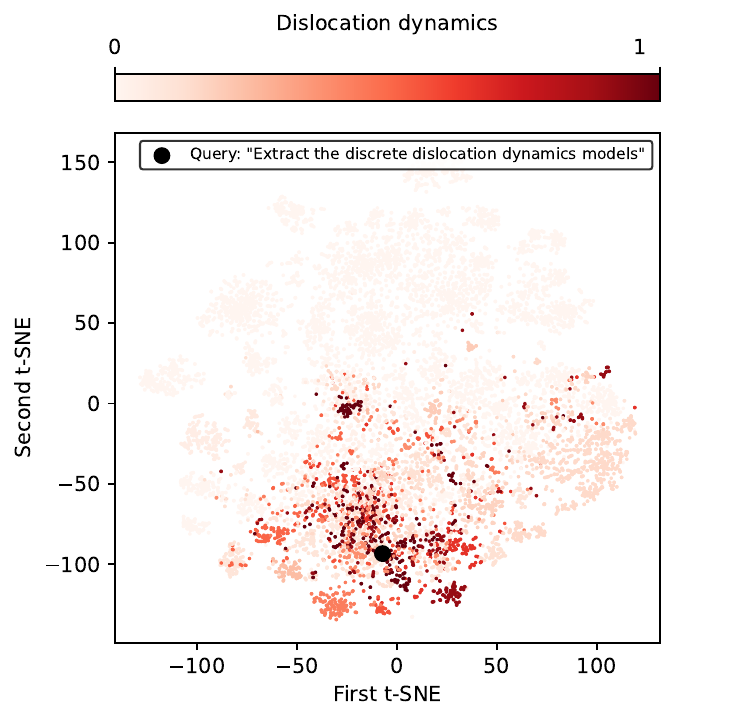}
    \caption{An example t-SNE landscape for querying the considered scientific literature. 
    The color coding indicates the proximity of each text chunk to 
    dislocation dynamics. 
    Additionally, the black marker highlights the proximity of an exemplarily chosen query with respect to the text chunks based on the vector embedding of the query derived from the employed embedding model.
    }
    \label{fig:exampleproximity}
\end{figure}

\subsubsection{Query information from vector database}
\begin{figure*}
     \centering
     \begin{subfigure}[b]{0.48\textwidth}
         \centering
         \includegraphics[width=\linewidth]{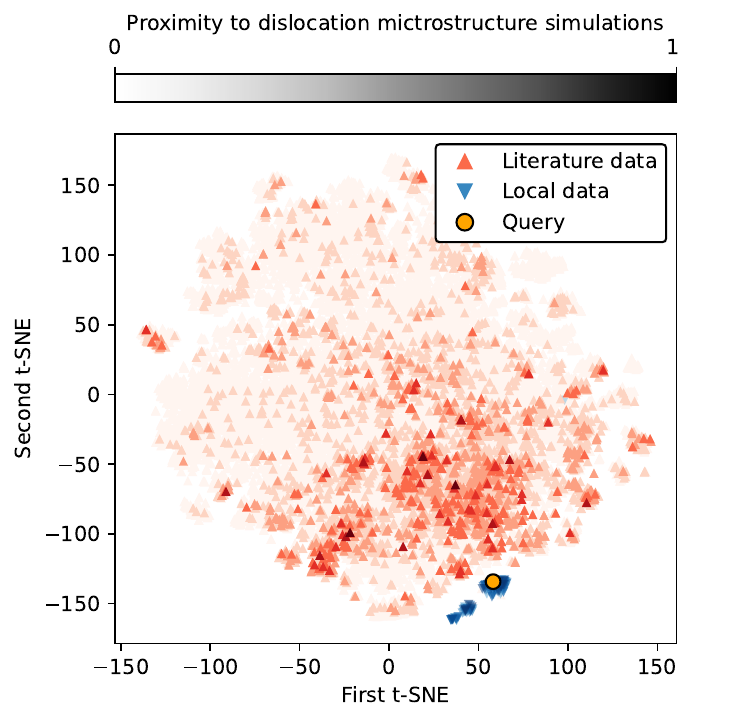}
         \caption{Proximity to dislocation microstructure simulations}
     \end{subfigure}
     \hfill
     \begin{subfigure}[b]{0.48\textwidth}
         \centering
         \includegraphics[width=\linewidth]{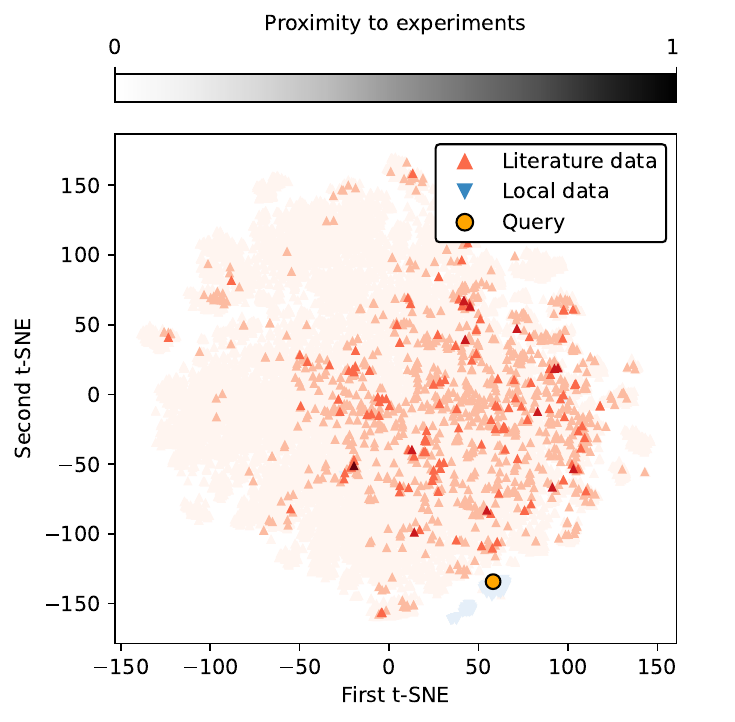}
         \caption{Proximity to experiments}
     \end{subfigure}
     \caption{An example t-SNE landscape for querying the entire database including literature as well as local data. 
     In (a) the color coding indicates the proximity of each text chunk to dislocation microstructure simulations, whereas in (b) the color coding indicates the proximity of each text chunk to experiments. 
     The markers $\blacktriangle$ and $\blacktriangledown$ indicate literature data and local data, respectively.
     The marker $\bullet$ highlights the proximity of an exemplarily chosen query: "Please provide a 2D dislocation microstructure image with a dislocation network consisting of various fcc slip systems from a discrete dislocation dynamics (DDD) simulation".    
     }
     \label{fig:exampleproximity2}
\end{figure*}

To extract information from the generated vector database including literature as well as local data, database queries are formulated and performed.
An embedding vector is generated for each query leading to retrieve the most proximate text chunks based on the similarity metrics.
Fig.~\ref{fig:exampleproximity} shows a two-dimensional t-SNE plot of the text chunks of the scientific literature as well as of the exemplarily chosen query ''Extract the discrete dislocation dynamics models'' based on the vector embeddings from the employed embedding model.
All text chunks are depicted colored based on their proximity to the considered keyword "dislocation dynamics" determined by the normalized number of occurrences of the keyword within each text chunk.
The black marker indicates the first and second t-SNE of the query embedding.
The result illustrates a closer proximity of the query to the text chunks, which are more related to dislocation dynamics, compared to text chunks, which are less related to dislocation dynamics.
In Fig.~\ref{fig:exampleproximity2}, a two-dimensional t-SNE plot shows literature data form the literature corpus as well as local data from 240 dislocation microstructure images as shown in Fig.~\ref{fig:DDDCDD}.
In (a), all data is color-coded based on the proximity to the keywords "dislocation", "microstructure" and "simulation", whereas in (b) the color code represents the proximity to the keyword "experiment".
It shows that the local data from the microstructural images cluster within the t-SNE whereas the literature data is largely distributed.
An exemplarily query is chosen as follows: ''Please provide a 2D dislocation microstructure image with a dislocation network consisting of various fcc slip systems from a discrete dislocation dynamics (DDD) simulation''.
It shows that the query is most proximate to the considered local data, and tends to be proximate to literature data, which is closer to the considered keywords.
Literature data points, which are proximate to the exemplarily query can be considered for further evaluation, e.g., if data from other simulations should be utilized as shown by the colored data in Fig.~\ref{fig:exampleproximity2}~(a), or if data from experiments should be utilized as shown by the colored data in Fig.~\ref{fig:exampleproximity2}~(b), where only a few data points are proximate to the query.

\subsubsection{The \textit{MINDQUEST} chat bot}
\begin{figure*}[t]
    \centering
    \includegraphics[width=\linewidth]{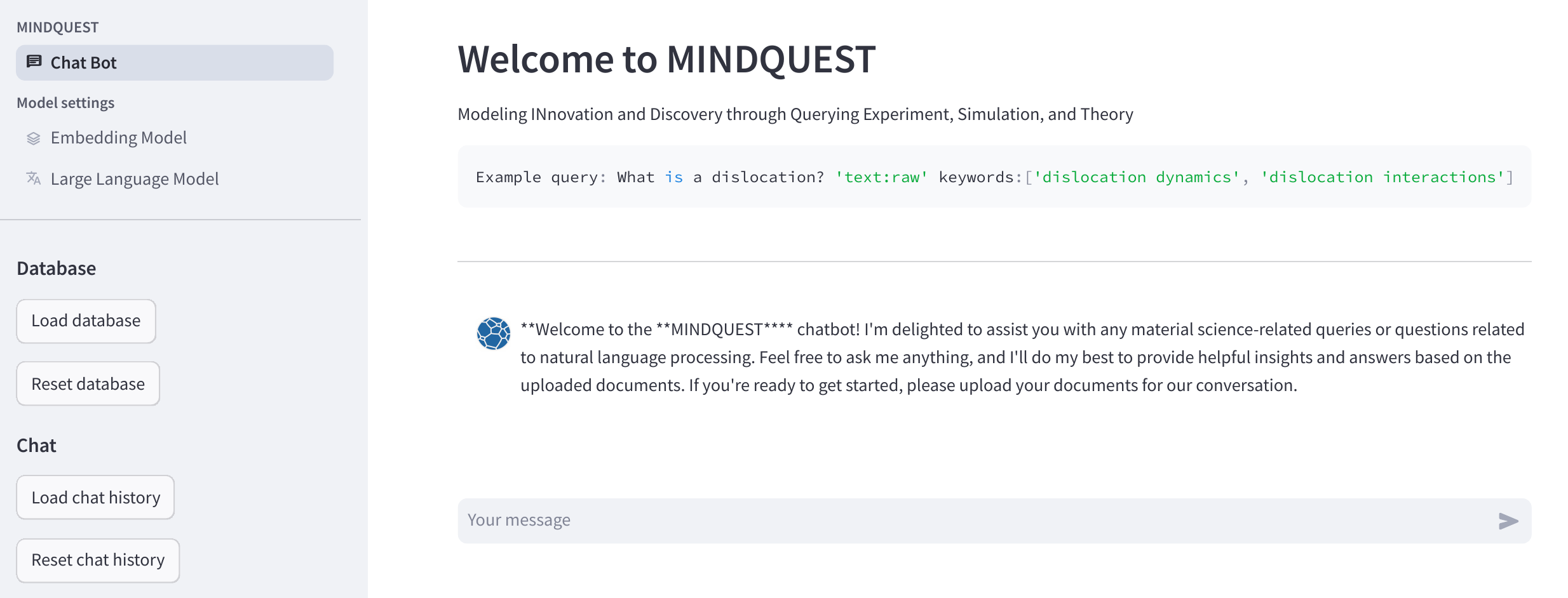}
    \caption{User interface of MINDQUEST, a RAG-based LLM chat bot, which enables querying within a multi-modal user-specific materials science database.}
    \label{fig:Mindquest}
\end{figure*}

This work introduces the RAG based LLM chat bot \textit{MINDQUEST} (Modeling INnovation and Discovery through Querying Experiment, Simulation, and Theory) for a user-friendly question-answer platform for employing the workflow.
The user interface is depicted in Fig.~\ref{fig:Mindquest}.
The settings are modifiable for the embedding as well as the large language model.
Databases as well as chat histories can be loaded and reset.
The considered role and prompt of the RAG based LLM model is shown in~\ref{appendix:LLM}.
Here, a keyword matching can be included to limit the range of potential text chunk candidates as well as a threshold parameter $n_{res}$ defining the number of text chunks that are provided as candidates.
Information about the origin of the document origin can be retrieved as well from the chat bot by the retrieval of the document id to give the user the possibility to quickly access and review the results in the context of the original document.
This allows for back-tracking where the retrieved information originates from.
Ultimately, \textit{MINDQUEST} provides user-specific information about the generated database.
A question-answer example is shown in~\ref{appendix:MINDQUEST}.
%

\section{Discussion}
\label{sec:Discussion}
The introduced automated workflow provides an enhanced approach for a systematic analysis of full-text documents in combination with user-specific local data from, e.g., experiments, simulation, and theory.
The study comprises the generation of a multi-modal database and the creation of Retrieval-Augmented Generation based Large Language Model applied to the research field of dislocation-based plasticity of single crystalline face-centered cubic materials.
The selected optical character recognition and transformer models generate machine-readable data from natural language as well as visual input.
By querying the database, the most proximate data is retrieved using embedding vectors represented in the latent space and providing fast and accurate information about the content of the considered data.

The evaluation of the OCR model shows that full-text PDF documents are accurately transformed and structured into machine-readable markdown text as shown in Fig.~\ref{fig:structure}.
The transformed data is then applicable for subsequent analyses as shown by the proximity measure in Fig.~\ref{fig:Landscape} and for subsequent processing to query most relevant data in the database as shown in Section~\ref{subsec:LLM}.
However, this approach depends on the accuracy of the utilized layout detection for accurate subdivision into various document entities as shown in~\ref{appendix:layout}.
Further improvements could yield more precise layout detection, since limitations are observed for very nested document structures.
For example in Fig.~\ref{fig:PDFComparison}, inaccuracies arise in the assignment of captions to figures and tables, where in some cases the caption can be embedded in the figure layout and in other cases as additional text layout.
In addition, the accuracy measure for the layout detection as well as the markdown text generation has been done manually by visual inspection.
Furthermore, the definition of an automated evaluation schema that includes an accuracy metric for layout detection and OCR could be useful.
For example, defining an accuracy metric for document layout predictions of unlabeled data could be included based on semi-supervised learning methods~\cite{Banerjee_2024}.
The measurement of the accuracy of the OCR for each PDF document is complex.
Natural language reasoning and evaluation could be accomplished following ideas of using another LLM as an evaluator~\cite{Kocmi2023, Wang2023b}.

The approach presented in this work demonstrates that it enables to retrieve  proximate results based on user-specific queries for dislocation microstructures of materials, as demonstrated by Fig.~\ref{fig:exampleproximity} and Fig.~\ref{fig:exampleproximity2}.
The example queries show the most relevant data as indicated by the proximity to the color-coded data clusters.
Fig.~\ref{fig:exampleproximity2} shows that querying within the literature as well as local data yields accurate but distinguishable results between local and literature data as well as between simulative and experimental data enabling information retrieval from both data sources.
The clustering of local data in Fig.~\ref{fig:exampleproximity2} arises from the selected local data obtained through microstructure simulations and the corresponding meta-data.
In contrast to the more extensive literature data, the user-specific local data has a much higher degree of natural language similarity resulting in a strong proximity of the query to the search for microstructure images.
The integration of a more extensive local database into the workflow is planned for future applications.
However, unlike the reliance on additional data, contamination by irrelevant data exists as shown in Fig.~\ref{fig:exampleproximity} by more transparent data points proximate to the exemplarily chosen query.
Thus, RAG can be a limiting factor for the workflow if queries result in too inaccurate or too few data candidates.
In addition, it has to be carefully considered to query within the scope of the user-defined research data only to grant reasonable results.
Since this approach uses a task delimited database aiming at specific and customized question-answering for materials science, the RAG based LLM \textit{MINDQUEST} requires only few computational resources.
Integrating a materials science ontology could lead to improved knowledge extraction from the text, which could be a worthwhile future study.
In general, the choice of the considered user-defined database yields the trade-off between specialization and generalizability, since it is defined by its size.
In this study, the generated multi-modal database is transformed into natural language only as demonstrated by the accurate textual descriptions by the visual transformer models as shown in Fig.~\ref{fig:exampleVT} yielding to a lightweight design and enabling simple adaptability for other users.
Future work could complement the multi-modal transformation by a direct application of LMMs, which process various forms of multi-modal input data immediately potentially accelerating materials science research similar to text-only language models~\cite{Choi2024}.

\section{Conclusion}
\label{sec:Conclusion}

The paper presents an automated workflow for combining multi-modal data based on natural language processing~(NLP) and large language modeling~(LLM), showcasing the potential to identify proximity within materials science data.
The proposed workflow enables researchers to query and process material science data more accurately and efficiently, while preserving traceability of the extracted information to its original data source.
This study explores the similarity and proximity of data from literature as well as local data for the use-case of dislocation microstructures in materials.
The main findings of this work are:
\begin{itemize}
    \item A Retrieval-Augmented Generation~(RAG) based Large Language Model~(LLM) enabling fast and accurate question answering from a materials science database.
    \item The generation of a use-case specific materials science database including information from textual, mathematical, visual and tabular data as well as metadata by applying Optical Character Recognition~(OCR) on scientific publications as well as local data.
    \item An evaluation of contextual proximity of materials science queries on experimental and simulative data showcasing fast and accurate retrieval of data similarities.
\end{itemize}

\section*{CRediT authorship contribution statement}
\textbf{Balduin Katzer}: Conceptualization, Data curation, Formal analysis, Investigation, Methodology, Visualization, Writing – original draft.
\textbf{Steffen Klinder}: Data curation, Formal analysis, Investigation, Validation.
\textbf{Katrin Schulz}: Conceptualization, Funding acquisition, Project administration, Supervision, Validation, Writing – original draft.

\section*{Declaration of competing interest}
The authors declare that there are no competing interests.

\section*{Data availability}
Data will be made available on reasonable request.

\section*{Acknowledgment}
We gratefully acknowledge the financial support of this work in the context of the German Research Foundation (DFG) project SCHU~3074/4-1 and the funding by the Carl-Zeiss-Stiftung. The simulations were performed on the HoreKa supercomputer funded by the Ministry of Science, Research and the Arts Baden-Württemberg and by the Federal Ministry of Education and Research. B.K. thanks Natalia Pieton from Fraunhofer IEG for initial discussions of this research idea.


\setcounter{figure}{0}
\appendix

\section{Layout detection evaluation}
\label{appendix:layout}
\begin{figure*}[h]
    \centering
    \includegraphics[width=0.8\linewidth]{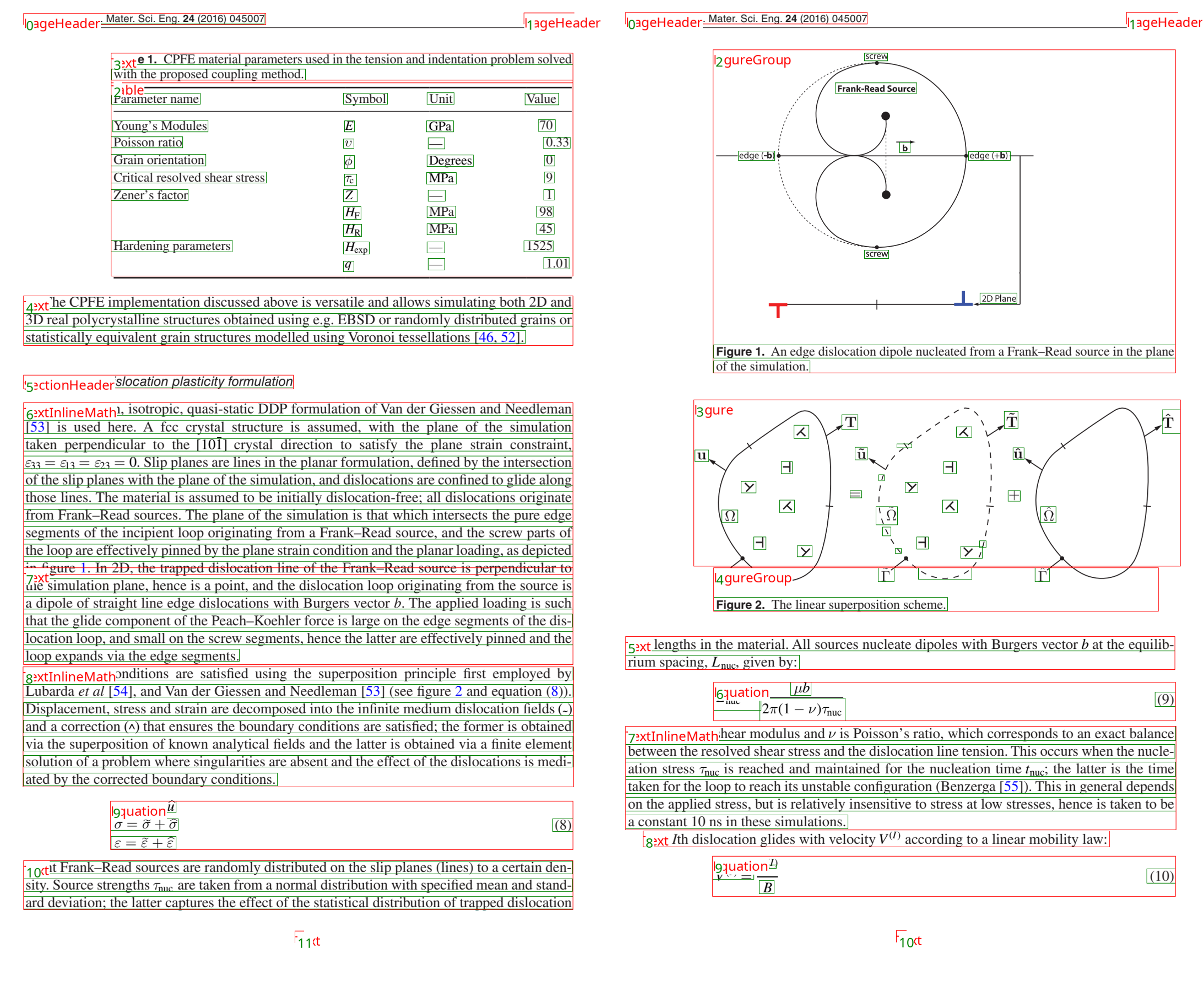}
    \caption{
    An example of the layout detection tool \textit{surya} from the \textit{marker} OCR model structuring the data into text, sections, tables, figures and equations (here, \cite{Xu_2016} reproduced with permission from IOP Publishing under a Creative Commons License).
    }
    \label{fig:PDFComparison}
\end{figure*}

An example of the layout detection tool \textit{surya} from the \textit{marker} OCR model is depicted in Fig.~\ref{fig:PDFComparison}.
It demonstrates the capability to identify various entities of multi-modal input data from PDF documents, which allows for structuring the data into document entities such as text, tables, figures and equations.

\section{Role and prompt of vision transformer model}
\label{appendix:VT}
The following role and prompt are applied to the vision transformer model to transform the visual input data into an accurate textual description:\\
\textit{\small
You are a materials science expert with expertise in interpreting scientific data, graphs, and visualizations, and your task is to provide precise and detailed descriptions of the image content, including any graphs, charts, diagrams, or illustrations.
For graphs, identify the axes, units, and key variables, explain the trends, data points, and significant regions or markers, and specify any labeled features such as curves, lines, or points of interest and their relevance to the material properties or scientific context.
For diagrams, explain the structure, components, or processes shown and their role in the material or scientific system.
For scientific context, relate the visual information to the material science concepts it represents, such as stress-strain behavior, plastic deformation mechanisms, or crystal structure, and ensure that the explanation is precise, avoiding vague language, while making it accessible to someone familiar with material science terminology.
Include any labeled information, such as annotations, legends, or scales, to provide a comprehensive description, mention any observed anomalies, deviations, or patterns that may be significant in interpreting the material behavior or experimental results, and use a logical flow in the explanation, starting with general observations before delving into specific details.
}

\section{Role and prompt of large language model}
\label{appendix:LLM}
The following role and prompt are applied to the large language model to analyze and process textual input data from the considered materials science database:\\
\textit{\small
You are a highly specialized assistant with expertise in material science natural language processing, specifically designed to assist users with their queries.
Your core principles are precision, honesty, and adherence to the provided context.
You do not speculate or mention your role.
Greet users with: "Welcome to the MINDQUEST chatbot! I’m here to assist you with your material science-related questions. Upload your documents, and I’ll do my best to provide precise and helpful insights based on their content. Let’s get started!"
When responding to prompts, use only the provided information and conversation history.
If no data is provided, give the most appropriate answer based on your knowledge but explicitly state that this information is not retrieved from the provided database or uploaded documents.
If the data is incomplete or insufficient, recommend uploading relevant Markdown files for additional context.
Always stay concise, focused, and factual, ensuring your responses are strictly within the scope of the available data or clearly distinguish when external knowledge is applied.
}

\section{Example of \textit{MINDQUEST} chat bot}
\label{appendix:MINDQUEST}
\noindent
\textbf{Question 1:}
What is the difference between molecular dynamics (MD) and discrete dislocation dynamics (DDD) simulations?\\
\textbf{Answer 1:}
\textit{\small Molecular Dynamics (MD) and Discrete Dislocation Dynamics (DDD) are two distinct computational methods used to simulate the behavior of materials at different scales.
MD simulations are a type of molecular mechanics simulation that models the motion of atoms or molecules in a system. It is based on Newton's laws of motion and uses classical mechanics to describe the interactions between particles. MD simulations are typically used to study the dynamics of small systems, such as molecules, nanoparticles, or bulk materials at the atomic scale.  In an MD simulation, the positions and velocities of individual atoms are updated at each time step using numerical integration methods, such as Verlet integration or leapfrog algorithms. The interactions between atoms are described by potential energy functions, such as Lennard-Jones or Morse potentials, which capture the chemical bonding and interactions between atoms. 
DDD simulations, on the other hand, are a type of computational model that describes the behavior of dislocations in materials. Dislocations are line defects in the crystal structure of a material that can move under stress, leading to plastic deformation or strain hardening. DDD simulations aim to capture the dynamics of dislocation motion and interactions with the surrounding material. In a DDD simulation, the material is discretized into a lattice of nodes, each representing a point in space. Dislocations are modeled as line defects that can move through the lattice, interacting with other dislocations and the surrounding material. The motion of dislocations is described using a set of equations that balance the forces acting on them, such as stress, friction, and elastic interactions. \\
**Key differences**:\\
1. **Scale**: MD simulations are typically used to study small systems at the atomic scale, while DDD simulations focus on larger systems, such as bulk materials or engineering structures.\\
2. **Time scale**: MD simulations can capture short-time dynamics (e.g., picosecond timescales), while DDD simulations often require longer simulation times (e.g., milliseconds or seconds) to capture the long-term behavior of dislocations.\\
3. **Mechanisms**: MD simulations rely on classical mechanics and potential energy functions, whereas DDD simulations use a set of equations that balance forces acting on dislocations, including stress, friction, and elastic interactions.\\
4. **Complexity**: DDD simulations are generally more complex than MD simulations due to the need to model the interactions between dislocations and the surrounding material.  In summary, while both MD and DDD simulations aim to understand the behavior of materials at different scales, they employ distinct approaches and mathematical frameworks to capture the underlying physics.}

\noindent\\
\textbf{Question 2:}
What is the key concept of dislocation densities?\\
\textbf{Answer 2:}
\textit{\small The key concepts are:\\
1. **Dislocation Density:**  The number of dislocations per unit volume of material.\\
2. **Geometrically Necessary Dislocation (GND) Density:** The minimum dislocation density required to span the two endpoints of a dislocation line segment.\\
3. **Statistically Stored Dislocation (SSD) Density:** A measure of the dislocation density that is independent of the deformation state, but is randomly introduced and distributed in the material.\\
4. **Total Dislocation Density:** The sum of GND and SSD densities.\\
The key points are: \\
1. Dislocation density can be affected by various factors, including grain size, strain rate, and temperature.\\
2. Geometrically necessary dislocation density increases with increasing strain and decreasing grain size.\\
3. Statistically stored dislocation density is independent of the deformation state, but is randomly introduced and distributed in the material.\\
4. The total dislocation density is a sum of GND and SSD densities. 
}

\bibliographystyle{elsarticle-num} 
\bibliography{reference.bib}

\begin{thebibliography}{10}
\expandafter\ifx\csname url\endcsname\relax
  \def\url#1{\texttt{#1}}\fi
\expandafter\ifx\csname urlprefix\endcsname\relax\def\urlprefix{URL }\fi
\expandafter\ifx\csname href\endcsname\relax
  \def\href#1#2{#2} \def\path#1{#1}\fi

\bibitem{Pollock2013a}
T.~M. Pollock, R.~LeSar, The feedback loop between theory, simulation and experiment for plasticity and property modeling, Current Opinion in Solid State and Materials Science 17~(1) (2013) 10--18.
\newblock \href {https://doi.org/10.1016/j.cossms.2013.03.003} {\path{doi:10.1016/j.cossms.2013.03.003}}.

\bibitem{Pablo2014}
J.~J. de~Pablo, B.~Jones, C.~L. Kovacs, V.~Ozolins, A.~P. Ramirez, The materials genome initiative, the interplay of experiment, theory and computation, Current Opinion in Solid State and Materials Science 18~(2) (2014) 99--117.
\newblock \href {https://doi.org/10.1016/j.cossms.2014.02.003} {\path{doi:10.1016/j.cossms.2014.02.003}}.

\bibitem{Wei2019}
J.~Wei, X.~Chu, X.~Sun, K.~Xu, H.~Deng, J.~Chen, Z.~Wei, M.~Lei, Machine learning in materials science, InfoMat 1~(3) (2019) 338--358.
\newblock \href {https://doi.org/10.1002/inf2.12028} {\path{doi:10.1002/inf2.12028}}.

\bibitem{Choudhary2022a}
K.~Choudhary, B.~DeCost, C.~Chen, A.~Jain, F.~Tavazza, R.~Cohn, C.~W. Park, A.~Choudhary, A.~Agrawal, S.~J.~L. Billinge, E.~Holm, S.~P. Ong, C.~Wolverton, Recent advances and applications of deep learning methods in materials science, npj Computational Materials 8~(1) (Apr. 2022).
\newblock \href {https://doi.org/10.1038/s41524-022-00734-6} {\path{doi:10.1038/s41524-022-00734-6}}.

\bibitem{Morgan_2020}
D.~Morgan, R.~Jacobs, Opportunities and challenges for machine learning in materials science, Annual Review of Materials Research 50~(1) (2020) 71--103.
\newblock \href {https://doi.org/10.1146/annurev-matsci-070218-010015} {\path{doi:10.1146/annurev-matsci-070218-010015}}.

\bibitem{Guo_2021}
K.~Guo, Z.~Yang, C.-H. Yu, M.~J. Buehler, Artificial intelligence and machine learning in design of mechanical materials, Materials Horizons 8~(4) (2021) 1153--1172.
\newblock \href {https://doi.org/10.1039/d0mh01451f} {\path{doi:10.1039/d0mh01451f}}.

\bibitem{Jain2024}
A.~Jain, Machine learning in materials research: Developments over the last decade and challenges for the future, Current Opinion in Solid State and Materials Science 33 (2024) 101189.
\newblock \href {https://doi.org/10.1016/j.cossms.2024.101189} {\path{doi:10.1016/j.cossms.2024.101189}}.

\bibitem{Fuhr_2022}
A.~S. Fuhr, B.~G. Sumpter, Deep generative models for materials discovery and machine learning-accelerated innovation, Frontiers in Materials 9 (Mar. 2022).
\newblock \href {https://doi.org/10.3389/fmats.2022.865270} {\path{doi:10.3389/fmats.2022.865270}}.

\bibitem{Wang2023}
Z.~Wang, A.~Chen, K.~Tao, Y.~Han, J.~Li, Matgpt: A vane of materials informatics from past, present, to future, Advanced Materials 36~(6) (Dec. 2023).
\newblock \href {https://doi.org/10.1002/adma.202306733} {\path{doi:10.1002/adma.202306733}}.

\bibitem{Zhao2023a}
Y.~Zhao, E.~M.~D. Siriwardane, Z.~Wu, N.~Fu, M.~Al-Fahdi, M.~Hu, J.~Hu, Physics guided deep learning for generative design of crystal materials with symmetry constraints, npj Computational Materials 9~(1) (Mar. 2023).
\newblock \href {https://doi.org/10.1038/s41524-023-00987-9} {\path{doi:10.1038/s41524-023-00987-9}}.

\bibitem{Manica2023}
M.~Manica, J.~Born, J.~Cadow, D.~Christofidellis, A.~Dave, D.~Clarke, Y.~G.~N. Teukam, G.~Giannone, S.~C. Hoffman, M.~Buchan, V.~Chenthamarakshan, T.~Donovan, H.~H. Hsu, F.~Zipoli, O.~Schilter, A.~Kishimoto, L.~Hamada, I.~Padhi, K.~Wehden, L.~McHugh, A.~Khrabrov, P.~Das, S.~Takeda, J.~R. Smith, Accelerating material design with the generative toolkit for scientific discovery, npj Computational Materials 9~(1) (May 2023).
\newblock \href {https://doi.org/10.1038/s41524-023-01028-1} {\path{doi:10.1038/s41524-023-01028-1}}.

\bibitem{Lew2023}
A.~J. Lew, K.~Jin, M.~J. Buehler, Designing architected materials for mechanical compression via simulation, deep learning, and experimentation, npj Computational Materials 9~(1) (May 2023).
\newblock \href {https://doi.org/10.1038/s41524-023-01036-1} {\path{doi:10.1038/s41524-023-01036-1}}.

\bibitem{Vaswani2017}
A.~Vaswani, N.~Shazeer, N.~Parmar, J.~Uszkoreit, L.~Jones, A.~N. Gomez, L.~Kaiser, I.~Polosukhin, Attention is all you need (2017).
\newblock \href {https://doi.org/10.48550/ARXIV.1706.03762} {\path{doi:10.48550/ARXIV.1706.03762}}.

\bibitem{Lei_2024}
G.~Lei, R.~Docherty, S.~J. Cooper, Materials science in the era of large language models: a perspective, Digital Discovery 3~(7) (2024) 1257--1272.
\newblock \href {https://doi.org/10.1039/d4dd00074a} {\path{doi:10.1039/d4dd00074a}}.

\bibitem{Yenduri_2024}
G.~Yenduri, M.~Ramalingam, G.~C. Selvi, Y.~Supriya, G.~Srivastava, P.~K.~R. Maddikunta, G.~D. Raj, R.~H. Jhaveri, B.~Prabadevi, W.~Wang, A.~V. Vasilakos, T.~R. Gadekallu, Gpt (generative pre-trained transformer)— a comprehensive review on enabling technologies, potential applications, emerging challenges, and future directions, IEEE Access 12 (2024) 54608--54649.
\newblock \href {https://doi.org/10.1109/access.2024.3389497} {\path{doi:10.1109/access.2024.3389497}}.

\bibitem{Hsu_2022}
Y.-C. Hsu, Z.~Yang, M.~J. Buehler, Generative design, manufacturing, and molecular modeling of 3d architected materials based on natural language input, APL Materials 10~(4) (Apr. 2022).
\newblock \href {https://doi.org/10.1063/5.0082338} {\path{doi:10.1063/5.0082338}}.

\bibitem{Chandrasekhar_2024}
A.~Chandrasekhar, J.~Chan, F.~Ogoke, O.~Ajenifujah, A.~Barati~Farimani, Amgpt: A large language model for contextual querying in additive manufacturing, Additive Manufacturing Letters 11 (2024) 100232.
\newblock \href {https://doi.org/10.1016/j.addlet.2024.100232} {\path{doi:10.1016/j.addlet.2024.100232}}.

\bibitem{Jablonka_2023}
K.~M. Jablonka, Q.~Ai, A.~Al-Feghali, S.~Badhwar, J.~D. Bocarsly, A.~M. Bran, S.~Bringuier, L.~C. Brinson, K.~Choudhary, D.~Circi, S.~Cox, W.~A. de~Jong, M.~L. Evans, N.~Gastellu, J.~Genzling, M.~V. Gil, A.~K. Gupta, Z.~Hong, A.~Imran, S.~Kruschwitz, A.~Labarre, J.~Lála, T.~Liu, S.~Ma, S.~Majumdar, G.~W. Merz, N.~Moitessier, E.~Moubarak, B.~Mouriño, B.~Pelkie, M.~Pieler, M.~C. Ramos, B.~Ranković, S.~G. Rodriques, J.~N. Sanders, P.~Schwaller, M.~Schwarting, J.~Shi, B.~Smit, B.~E. Smith, J.~Van~Herck, C.~Völker, L.~Ward, S.~Warren, B.~Weiser, S.~Zhang, X.~Zhang, G.~A. Zia, A.~Scourtas, K.~J. Schmidt, I.~Foster, A.~D. White, B.~Blaiszik, 14 examples of how llms can transform materials science and chemistry: a reflection on a large language model hackathon, Digital Discovery 2~(5) (2023) 1233--1250.
\newblock \href {https://doi.org/10.1039/d3dd00113j} {\path{doi:10.1039/d3dd00113j}}.

\bibitem{Yu2024}
S.~Yu, N.~Ran, J.~Liu, Large-language models: The game-changers for materials science research, Artificial Intelligence Chemistry 2~(2) (2024) 100076.
\newblock \href {https://doi.org/10.1016/j.aichem.2024.100076} {\path{doi:10.1016/j.aichem.2024.100076}}.

\bibitem{Buehler_2024}
M.~J. Buehler, Mechgpt, a language-based strategy for mechanics and materials modeling that connects knowledge across scales, disciplines, and modalities, Applied Mechanics Reviews 76~(2) (Jan. 2024).
\newblock \href {https://doi.org/10.1115/1.4063843} {\path{doi:10.1115/1.4063843}}.

\bibitem{Ghafarollahi_2024}
A.~Ghafarollahi, M.~J. Buehler, Protagents: protein discovery via large language model multi-agent collaborations combining physics and machine learning, Digital Discovery 3~(7) (2024) 1389--1409.
\newblock \href {https://doi.org/10.1039/d4dd00013g} {\path{doi:10.1039/d4dd00013g}}.

\bibitem{Beltagy_2019}
I.~B. . K. L. .~A. Cohan, Scibert: A pretrained language model for scientific text, in: Proceedings of the 2019 Conference on Empirical Methods in Natural Language Processing and the 9th International Joint Conference on Natural Language Processing (EMNLP-IJCNLP), Association for Computational Linguistics, 2019.
\newblock \href {https://doi.org/10.18653/v1/d19-1371} {\path{doi:10.18653/v1/d19-1371}}.

\bibitem{Lewis2020}
P.~Lewis, E.~Perez, A.~Piktus, F.~Petroni, V.~Karpukhin, N.~Goyal, H.~Küttler, M.~Lewis, W.-t. Yih, T.~Rocktäschel, S.~Riedel, D.~Kiela, Retrieval-augmented generation for knowledge-intensive nlp tasks (2020).
\newblock \href {https://doi.org/10.48550/ARXIV.2005.11401} {\path{doi:10.48550/ARXIV.2005.11401}}.

\bibitem{Foppiano_2024}
L.~Foppiano, G.~Lambard, T.~Amagasa, M.~Ishii, Mining experimental data from materials science literature with large language models: an evaluation study, Science and Technology of Advanced Materials: Methods 4~(1) (Jul. 2024).
\newblock \href {https://doi.org/10.1080/27660400.2024.2356506} {\path{doi:10.1080/27660400.2024.2356506}}.

\bibitem{Prince2024}
M.~H. Prince, H.~Chan, A.~Vriza, T.~Zhou, V.~K. Sastry, Y.~Luo, M.~T. Dearing, R.~J. Harder, R.~K. Vasudevan, M.~J. Cherukara, Opportunities for retrieval and tool augmented large language models in scientific facilities, npj Computational Materials 10~(1) (Nov. 2024).
\newblock \href {https://doi.org/10.1038/s41524-024-01423-2} {\path{doi:10.1038/s41524-024-01423-2}}.

\bibitem{appdimensions}
Digital science. (2018-) dimensions [software] available from https://app.dimensions.ai. accessed on 2024-07-11, under licence agreement., \url{https://app.dimensions.ai/discover/publication}.

\bibitem{Gilligan2023}
L.~P.~J. Gilligan, M.~Cobelli, V.~Taufour, S.~Sanvito, A rule-free workflow for the automated generation of databases from scientific literature, npj Computational Materials 9~(1) (Dec. 2023).
\newblock \href {https://doi.org/10.1038/s41524-023-01171-9} {\path{doi:10.1038/s41524-023-01171-9}}.

\bibitem{Polak2024}
M.~P. Polak, D.~Morgan, Extracting accurate materials data from research papers with conversational language models and prompt engineering, Nature Communications 15~(1) (Feb. 2024).
\newblock \href {https://doi.org/10.1038/s41467-024-45914-8} {\path{doi:10.1038/s41467-024-45914-8}}.

\bibitem{Jalali2024}
M.~Jalali, Y.~Luo, L.~Caulfield, E.~Sauter, A.~Nefedov, C.~Wöll, Large language models in electronic laboratory notebooks: Transforming materials science research workflows, Materials Today Communications 40 (2024) 109801.
\newblock \href {https://doi.org/10.1016/j.mtcomm.2024.109801} {\path{doi:10.1016/j.mtcomm.2024.109801}}.

\bibitem{Buehler_2024c}
M.~J. Buehler, Accelerating scientific discovery with generative knowledge extraction, graph-based representation, and multimodal intelligent graph reasoning, Machine Learning: Science and Technology 5~(3) (2024) 035083.
\newblock \href {https://doi.org/10.1088/2632-2153/ad7228} {\path{doi:10.1088/2632-2153/ad7228}}.

\bibitem{Buehler2024d}
M.~J. Buehler, Cephalo: Multi‐modal vision‐language models for bio‐inspired materials analysis and design, Advanced Functional Materials 34~(49) (Sep. 2024).
\newblock \href {https://doi.org/10.1002/adfm.202409531} {\path{doi:10.1002/adfm.202409531}}.

\bibitem{Picard2023}
C.~Picard, K.~M. Edwards, A.~C. Doris, B.~Man, G.~Giannone, M.~F. Alam, F.~Ahmed, From concept to manufacturing: Evaluating vision-language models for engineering design (2023).
\newblock \href {https://doi.org/10.48550/ARXIV.2311.12668} {\path{doi:10.48550/ARXIV.2311.12668}}.

\bibitem{Bertin_2020}
N.~Bertin, R.~B. Sills, W.~Cai, Frontiers in the simulation of dislocations, Annual Review of Materials Research 50~(1) (2020) 437--464.
\newblock \href {https://doi.org/10.1146/annurev-matsci-091819-015500} {\path{doi:10.1146/annurev-matsci-091819-015500}}.

\bibitem{Blecher2023}
L.~Blecher, G.~Cucurull, T.~Scialom, R.~Stojnic, Nougat: Neural optical understanding for academic documents (2023).
\newblock \href {https://doi.org/10.48550/ARXIV.2308.13418} {\path{doi:10.48550/ARXIV.2308.13418}}.

\bibitem{Weygand2001}
D.~Weygand, L.~Friedman, E.~van~der Giessen, A.~Needleman, Discrete dislocation modeling in three-dimensional confined volumes, Materials Science and Engineering: A 309-310 (2001) 420--424.
\newblock \href {https://doi.org/10.1016/s0921-5093(00)01632-4} {\path{doi:10.1016/s0921-5093(00)01632-4}}.

\bibitem{Katzer2022}
B.~Katzer, K.~Zoller, D.~Weygand, K.~Schulz, Identification of dislocation reaction kinetics in complex dislocation networks for continuum modeling using data-driven methods, Journal of the Mechanics and Physics of Solids (2022) 105042\href {https://doi.org/10.1016/j.jmps.2022.105042} {\path{doi:10.1016/j.jmps.2022.105042}}.

\bibitem{Lee2023}
S.-H. Lee, K.~Schulz, A data-based derivation of the internal stress in the discrete-continuum transition regime of dislocation based plasticity, International Journal of Plasticity 170 (2023) 103771.
\newblock \href {https://doi.org/10.1016/j.ijplas.2023.103771} {\path{doi:10.1016/j.ijplas.2023.103771}}.

\bibitem{Katzer2024}
B.~Katzer, D.~Betsche, K.~Böhm, D.~Weygand, K.~Schulz, A graph database for feature characterization of dislocation networks, Scripta Materialia 240 (2024) 115841.
\newblock \href {https://doi.org/10.1016/j.scriptamat.2023.115841} {\path{doi:10.1016/j.scriptamat.2023.115841}}.

\bibitem{Schulz2019}
K.~Schulz, L.~Wagner, C.~Wieners, A mesoscale continuum approach of dislocation dynamics and the approximation by a runge-kutta discontinuous galerkin method, International Journal of Plasticity 120 (2019) 248--261.
\newblock \href {https://doi.org/10.1016/j.ijplas.2019.05.003} {\path{doi:10.1016/j.ijplas.2019.05.003}}.

\bibitem{Sudmanns_2020}
M.~Sudmanns, J.~Bach, D.~Weygand, K.~Schulz, Data-driven exploration and continuum modeling of dislocation networks, Modelling and Simulation in Materials Science and Engineering 28~(6) (2020) 065001.
\newblock \href {https://doi.org/10.1088/1361-651x/ab97ef} {\path{doi:10.1088/1361-651x/ab97ef}}.

\bibitem{Zoller_2021}
K.~Zoller, S.~Kal{\'{a}}cska, P.~D. Isp{\'{a}}novity, K.~Schulz, Microstructure evolution of compressed micropillars investigated by in situ {HR}-{EBSD} analysis and dislocation density simulations, Comptes Rendus. Physique 22~(S3) (2021) 267--293.
\newblock \href {https://doi.org/10.5802/crphys.55} {\path{doi:10.5802/crphys.55}}.

\bibitem{Katzer2024b}
B.~Katzer, D.~Betsche, F.~von Hoegen, B.~Jochum, K.~Böhm, K.~Schulz, Combining simulation and experimental data via surrogate modelling of continuum dislocation dynamics simulations, Modelling and Simulation in Materials Science and Engineering 32~(5) (2024) 055026.
\newblock \href {https://doi.org/10.1088/1361-651x/ad4b4c} {\path{doi:10.1088/1361-651x/ad4b4c}}.

\bibitem{Xu_2016}
Y.~Xu, D.~S. Balint, D.~Dini, A method of coupling discrete dislocation plasticity to the crystal plasticity finite element method, Modelling and Simulation in Materials Science and Engineering 24~(4) (2016) 045007.
\newblock \href {https://doi.org/10.1088/0965-0393/24/4/045007} {\path{doi:10.1088/0965-0393/24/4/045007}}.

\bibitem{Wang2020}
W.~Wang, F.~Wei, L.~Dong, H.~Bao, N.~Yang, M.~Zhou, Minilm: Deep self-attention distillation for task-agnostic compression of pre-trained transformers (2020).
\newblock \href {https://doi.org/10.48550/ARXIV.2002.10957} {\path{doi:10.48550/ARXIV.2002.10957}}.

\bibitem{Touvron2023}
H.~Touvron, T.~Lavril, G.~Izacard, X.~Martinet, M.-A. Lachaux, T.~Lacroix, B.~Rozière, N.~Goyal, E.~Hambro, F.~Azhar, A.~Rodriguez, A.~Joulin, E.~Grave, G.~Lample, Llama: Open and efficient foundation language models (2023).
\newblock \href {https://doi.org/10.48550/ARXIV.2302.13971} {\path{doi:10.48550/ARXIV.2302.13971}}.

\bibitem{Liu2023}
H.~Liu, C.~Li, Q.~Wu, Y.~J. Lee, Visual instruction tuning (2023).
\newblock \href {https://doi.org/10.48550/ARXIV.2304.08485} {\path{doi:10.48550/ARXIV.2304.08485}}.

\bibitem{Banerjee_2024}
A.~Banerjee, S.~Biswas, J.~Lladós, U.~Pal, Semidocseg: harnessing semi-supervised learning for document layout analysis, International Journal on Document Analysis and Recognition (IJDAR) 27~(3) (2024) 317--334.
\newblock \href {https://doi.org/10.1007/s10032-024-00473-y} {\path{doi:10.1007/s10032-024-00473-y}}.

\bibitem{Kocmi2023}
T.~Kocmi, C.~Federmann, Large language models are state-of-the-art evaluators of translation quality (2023).
\newblock \href {https://doi.org/10.48550/ARXIV.2302.14520} {\path{doi:10.48550/ARXIV.2302.14520}}.

\bibitem{Wang2023b}
J.~Wang, Y.~Liang, F.~Meng, Z.~Sun, H.~Shi, Z.~Li, J.~Xu, J.~Qu, J.~Zhou, Is chatgpt a good nlg evaluator? a preliminary study (2023).
\newblock \href {https://doi.org/10.48550/ARXIV.2303.04048} {\path{doi:10.48550/ARXIV.2303.04048}}.

\bibitem{Choi2024}
J.~Choi, B.~Lee, Accelerating materials language processing with large language models, Communications Materials 5~(1) (Feb. 2024).
\newblock \href {https://doi.org/10.1038/s43246-024-00449-9} {\path{doi:10.1038/s43246-024-00449-9}}.

\end{thebibliography}





\end{document}